\documentclass[a4paper,aps,prl,superscriptaddress,reprint,longbibliography]{revtex4-2}

\usepackage[utf8]{inputenc}
\usepackage[english]{babel}
\usepackage{amsmath,amssymb}
\usepackage{graphicx}
\usepackage{color}
\usepackage{bm}
\usepackage{physics}
\usepackage{upgreek}
\usepackage{setspace}

\begin{document}
	
\title{Axion Dark Matter Modulated Spin Wave Interferometry}

\author{Jiayi Liu}
\affiliation{School of Physics, Beijing Institute of Technology, Beijing, 100081, China}

	\author{Chen-Hui Xie}
\affiliation{School of Physics, Beijing Institute of Technology, Beijing, 100081, China}

\author{Runyu Lei}
\affiliation{School of Physics and Astronomy, Beijing Normal University, Beijing 100875, China.}

\author{Sichun Sun}
\email{sichunssun@bit.edu.cn}
\affiliation{School of Physics, Beijing Institute of Technology, Beijing, 100081, China}

\author{Yu Gao}
\email{gaoyu@ihep.ac.cn}
\affiliation{Institute of High Energy Physics, Chinese Academy of Sciences, Beijing 100049, China}
	\author{Jinxing Zhang}
\email{jxzhang@mail.bnu.edu.cn}
\affiliation{School of Physics and Astronomy, Beijing Normal University, Beijing 100875, China.}

\begin{abstract}
We propose a novel asymmetric 
spin wave interferometer to detect ultralight axion dark matter. The axion modulates spin-wave properties via a weak effective magnetic field in ferromagnets. The interferometer splits a spin-wave source into two paths of different lengths and sets them to interfere destructively. The system then converts the axion-induced phase shift into a measurable magnetization oscillation that can radiate electromagnetic waves and generate electrical signals via Faraday induction. The signal-to-noise ratios have been evaluated for three detection schemes: the linear amplifier, the single-photon detector, and the electrical signal detection approach, accounting for both magnetization fluctuation and thermal noise. The accessible axion mass range is approximately $10^{-8}$ eV to $10^{-6}$ eV, set by the spin wave propagation length and the relaxation time.
\end{abstract}
\maketitle
\textbf{INTRODUCTION} \par
The axion, a hypothetical particle originally proposed by Peccei and Quinn (PQ) in 1977, emerges as a compelling solution to the strong CP problem in quantum chromodynamics (QCD) \cite{Peccei:1977hh,Peccei:1977ur,Weinberg:1977ma,Wilczek:1977pj}.
Beyond its role in solving the strong CP problem, the axion has gained prominence as a leading candidate for light dark matter. Its extremely weak coupling to Standard Model particles and low mass (typically  \(10^{-6}-10^{-3}\) eV) make it a viable component of the cold dark matter halo. Axion-like particles (ALPs) are hypothetical pseudoscalar bosons that arise in a variety of extensions of the Standard Model and share many phenomenological properties with the QCD axion. Unlike the QCD axion, ALPs can exhibit a broad range of masses ($10^{-13}-10^{-2}$ eV) \cite{Sikivie:2020zpn}. Hereafter, "axions" refers collectively to QCD axions and ALPs. Their extremely weak interactions with ordinary matter make experimental detection highly challenging, motivating the development of novel search strategies.

Direct detection of axions has been a focal point of experimental physics, with approaches spanning a wide array of techniques \cite{Millar:2016cjp,Mitridate:2020kly,PhysRevD.101.096013,Budker:2013hfa,PhysRevLett.108.161803,Lee:2024toz,Kalia:2024eml,Sikivie:2009qn,PhysRevLett.58.1799,PhysRevD.79.107301,Gramolin:2020ict,Reina-Valero:2024wqx,Diehl_2023,https://doi.org/10.1155/2017/6432354,PhysRevA.97.042506,Berlin_2021,Chao2023liu,Duan2022nuy,Chen:2020cbs,Gao:2022zxc,dror2025axionproductiondetectionusing,chen2026searchqcdaxiondark,gao2026searchingaxiondarkmatter}. Cavity-based experiments, such as ADMX, exploit the axion photon coupling \(g_{a\gamma\gamma}\) to convert axions into microwave photons in strong magnetic fields \cite{Sikivie:1983ip,ADMX:2025vom,Irastorza:2018dyq,ParticleDataGroup:2018ovx,Lawson:2019brd,Sikivie:2020zpn,Rybka:2014xca,Rettaroli_2024,PhysRevD.99.101101}. 
In recent years, condensed matter systems have emerged as a promising platform for axion detection, leveraging their unique quantum and collective phenomena \cite{Lei:2025vek,Rivera2009ASR,Fiebig2009CurrentTO,10.21468/SciPostPhys.6.4.046,article,HEHL20081141,PhysRevLett.132.181801}. The CASPEr experiment exploits the axion nucleon coupling to induce an oscillating nuclear electric dipole moment (EDM), which resonantly drives nuclear spin precession when its frequency matches the nuclear spin precession frequency, thereby amplifying the signal\cite{Budker:2013hfa,Garcon_2017,Graham:2013gfa,Blanchard2015,WANG201827,dror2025axionproductiondetectionusing}. In contrast, the QUAX-ae experiment equates the axion electron coupling to an oscillating magnetic field, which resonantly amplifies the signal when its frequency matches the Larmor frequency of the material\cite{BARBIERI2017135,PhysRevD.101.096013,Crescini_2018,PhysRevLett.124.171801,Berlin_2024}. 

Axions can couple to fermions, such as nucleons and electrons, generating an effective oscillating magnetic field acting on spins. Unlike previous axion detection experiments that rely on resonant spin precession, our approach exploits the propagation properties of spin waves, in which the phase evolution can be modulated by the axion-induced effective magnetic field. Spin waves exhibit advantages of low propagation loss and long propagation distance\cite{94ead522f2ea4bb5876d146b6ba6b66f,Karenowska2014,Flebus_2024}. Their intrinsic frequencies span a broad range from GHz to THz, and parameters such as amplitude and phase can be efficiently manipulated by various physical fields \cite{9706176,Flebus_2024}. The spin wave interferometer uses this method to control the 0 and 1 states of spin wave logic gates \cite{10.1063/1.2975235}. Moreover, the fabrication processes for magnetic thin films and nanostructures that host spin waves are compatible with existing semiconductor technology, thereby laying a solid foundation for scalable and economical integration \cite {Barman_2021}. 

The axion-induced phase difference is amplified into a detectable magnetization oscillation signal by designing a spin wave interferometer. This oscillation radiates electromagnetic waves that can be detected by a linear amplifier or a single-photon detector. It is worth noting that Balynsky et al. \cite{Balynsky:2016uvn} have experimentally demonstrated a magnetometer based on a Mach-Zehnder spin wave interferometer and Faraday's law of induction, confirming the sensitive response of such a structure to magnetic field changes. Their work showed that the interferometer is most sensitive to phase variations when the spin waves are set to destructive interference, and they measured an output voltage of 10$\,\mu$V at the destructive point at room temperature. 
 Our approach leverages well-established techniques of ferromagnetic resonance and spin-wave interferometry, offering a complementary and experimentally feasible new pathway for axion dark matter detection.

\textbf{MODEL SETUP} \par
The axion field  $a$ is modeled as :
\begin{align}
	a(\bm{r},t) &= a_0 \cos(m_a\bm{v}_a\cdot\bm{r}-\omega_a t),
\end{align}
 where $v_a\sim 10^{-3}\,c$ is the solar system's relative velocity to the Galactic dark matter halo,  $\omega_a= m_a(1+\frac{1}{2}v_a^2)\approx m_a$, and $m_a$ is the axion mass. The amplitude \(a_0\) is determined by the local dark matter density, given by $\rho_{\text{DM}}=\frac{1}{2}m_a^2|a_0|^2$, where $\rho_{\text{DM}}$ is the local DM density and $\rho_{\text{DM}}\approx 0.4~ \text{GeV}/ \text{cm}^{3}$ \cite{Millar:2016cjp}. Based on the Hamiltonian describing the axion-electron coupling, the axion effective magnetic field is expressed as:

\begin{align}	
	\begin{split}
		\bm{B}_{\mathrm{eff}}(\boldsymbol{r},t)
		&= -\frac{\mu_0 g_{ae}}{2 m_e \gamma} \nabla a(\boldsymbol{r},t) \\
		&= \frac{\mu_0 g_{ae} a_0}{2 m_e \gamma} m_a\bm{v_a}
		\sin\bigl( m_a\bm{v_a}\cdot \boldsymbol{r} - \omega_a t \bigr).
	\end{split}
\end{align}

Here, we investigate the magnetization dynamics in a uniaxial ferromagnetic system with energy dissipation, governed by the Landau-Lifshitz-Gilbert (LLG) equation:  
\begin{equation} \label{eq:LLG_modified}  
	\frac{d\bm M}{dt}  = -\gamma \bm M \times \bm H + \frac{\alpha}{M_s} \bm M \times \frac{d\bm M}{dt},  
\end{equation}  
where $\gamma= 2.21 \times 10^{5} \ \mathrm{rad/(s\cdot (A/m))}$ represents the gyromagnetic ratio, $\bm{M}$ is the magnetization. 
 $\bm{H}$ denotes the effective magnetic field, which includes not only the externally applied magnetic field but also various effective magnetic fields generated inside the ferromagnet, such as the magnetocrystalline anisotropy field, the demagnetizing field, the exchange field, and the axion effective magnetic field, among others. 
 The parameter $\alpha$ is the Gilbert damping constant, a positive dimensionless quantity that phenomenologically characterizes energy dissipation during the precession of magnetization, with typical values ranging from approximately $10^{-5}$ to $10^{-1}$ \cite{1353448,Magneticdamping,9706176,hauser_yttrium_2016}. $M_s$ is the saturation magnetization. 

The axion effective magnetic field influences the spin wave dispersion. According to the Landau–Lifshitz theory, the dispersion relation of a ferromagnet is given by:
\begin{equation}  
	   \Omega(k, \boldsymbol{r}, t)
	= \gamma \bigl(H_0 + H_{\mathrm{eff},z}(\boldsymbol{r}, t) \bigr) + D k^2=\omega(k)+\delta \omega,
\end{equation}  
where $\omega(k)=\gamma H_0 + D k^2$ is the spin wave eigenfrequency \cite{Stancil2009SpinWT} and $\delta \omega$ denotes the axion-induced frequency shift. $H_0$ includes constant terms in the z-direction, such as the external static magnetic field and the effective magnetic fields generated inside the ferromagnetic material.  $D$ is the exchange stiffness constant. 
 $\bm{H}_{\mathrm{eff}}(\boldsymbol{r},t)
= \frac{\bm{B}_{\mathrm{eff}}(\boldsymbol{r},t)}{\mu_0}$ is axion effective magnetic intensity.
Since the spin wave frequency is sensitive only to the field component along the magnetization equilibrium axis $z$, the axion-induced frequency shift is determined solely by the $z$-projection of the axion effective field:
\begin{align}
|\delta \omega|          =\gamma |H_{\mathrm{eff},z}|            =3.7 \times 10^{-10}\big(\frac{g_{ae}}{10^{-13}}\big)\,\text{rad/s}.
\end{align}

In spintronics, ferromagnetic resonance is commonly used to probe the resonance frequency of a ferromagnet \cite{Ferromagneticresonance}. A small transverse alternating magnetic field $\bm{h}$ with frequency $\omega_r$ is applied to the ferromagnetic material, and resonance occurs when the frequency of this alternating field matches the frequency of the uniform precession mode $\omega(k=0)=\gamma H_0=\omega_r$. In ferromagnetic resonance studies, right-handed polarizations of magnetization $m_+=m_x+im_y $ and magnetic field  $h_+=h_x+ih_y $ are often adopted \cite{Gurevich1996MagnetizationOA}. The right-handed  magnetic susceptibility of the ferromagnet is given by:
\begin{align}
	\chi_+
	= \frac{\gamma M_0}{\Omega(t,k) - \omega_r + i\alpha \omega_r}\label{eq:chi},
\end{align}
In this case, $\Omega(t,k=0)=\gamma (H_0+\frac{g_{ae}a_0}{2m_{e}\gamma}m_a v_{az} \sin \omega_a t)=\omega_r  +|\delta \omega| \sin \omega_a t$. The phase of the magnetic susceptibility can be expressed as:
\begin{align}
	\phi_\chi(t)=\arctan\left(\frac{\alpha \omega_r}{\Omega(t)-\omega_r}\right)=\arctan\left(\frac{\alpha \omega_r}{|\delta \omega| \sin \omega_a t }\right),
\end{align}
for $|\delta \omega|\ll\alpha\omega_r$, $\phi_\chi(t)
\approx \frac{\pi}{2} - \frac{|\delta \omega| \sin\omega_a t}{\alpha \omega_r}=\frac{\pi}{2} -\frac{g_{ae}a_0}{2m_{e}\alpha \omega_r}m_a v_{az} \sin \omega_a t$. This phase can be regarded as the phase difference between \(h_+\) and \(m_+\). The phase change induced by the axion effective magnetic field is given by:
\begin{align}
|\delta \phi_\chi|  &=\frac{g_{ae}a_0}{2m_{e}\alpha \omega_r}m_a v_{az} \nonumber\\
&=7.4\times 10^{-16}\big(\frac{g_{ae}}{10^{-13}}\big)\big(\frac{5\,\text{GHz}}{\omega_r}\big)\big(\frac{10^{-4}}{\alpha}\big).
\end{align}

\textbf{AXION-INDUCED MAGNETIZATION} \par
We now extend the discussion from uniform precession at $k =  0$ to propagating spin waves with $k \neq 0$, $\omega = \gamma H_0+Dk^2$. Owing to the influence of the axion magnetic field on the spin wave frequency, the accumulation of frequency variation over time during spin wave propagation leads to a change in the spin wave phase:
\begin{align}
	\delta \phi &=-\int_{0}^{t} \delta \omega dt'= -\gamma \int_{0}^{t}B_\mathrm{eff}(\bm{r}, t')_z dt'\nonumber\\
	&=\frac{g_{ae}a_0}{2m_{e}}v_{az}\cos(\omega_a t-\phi_a)
\end{align}
As the axion evolves with time, the phase of the spin wave also varies at the frequency $\omega_a$.
The amplitude of this modulation is given by:
\begin{align}
	|\delta \phi|  =2.4\times 10^{-19}\big(\frac{g_{ae}}{10^{-13}}\big)\big(\frac{10^{-6}\text{eV}}{m_a}\big)
\end{align}

Interferometry is a common technique for measuring phase changes, as it can significantly amplify signals and enhance sensitivity. Splitting a spin wave into two waves that travel along different paths and then recombining them produces a phase difference between the two paths that causes a change in the combined amplitude. Spin waves originating from the same source, after propagating through materials of different lengths $L_1$ and $L_2$, recombine at time $t$. The two spin waves are expressed as
\begin{align}
	\psi_1(t) &=A \, e^{i\phi_{1}}= A \, e^{i(kL_1 - \omega t + \delta\phi_{1})}, \\
	\psi_2(t) &=A \, e^{i\phi_{2}}= A \, e^{i(kL_2 - \omega t + \delta\phi_{2})}.
\end{align}
where $\delta\phi_{1}$ and $\delta\phi_{2}$ are the phase difference induced by the axion.
$A=\chi_+h_+$ is the complex amplitude of the spin wave. The superposition of the two spin waves is given by:
\begin{align}
	\psi(t) &= \psi_1(t) + \psi_2(t) \nonumber \\
	&= A e^{-i\omega t} \left( e^{i(kL_1+ \delta\phi_{1})} + e^{i(kL_2+\delta\phi_{2})} \right).
\end{align}  
 
 The modulus of the resultant complex amplitude $\tilde{A}$ is computed as:
 \begin{align}
 |\tilde{A}| =  2|A| \left|\cos\frac{\theta_2-\theta_1}{2}\right|. \label{eq:amplitude}
\end{align}
 where $\theta_1 = kL_1+\delta\phi_{1}, \ \theta_2 = kL_2+\delta\phi_{2}$. Different path lengths lead to different times for the axion-induced frequency shift to accumulate, producing a phase difference:
 \begin{align}
 	&\Delta \theta(L_1,L_2)= \theta_2-\theta_1\nonumber\\
 	&=k\Delta L-\frac{g_{ae}a_0}{m_{e}}v_{az}\sin(\omega_a t-\frac{\omega_a (L_1+L_2)}{2v_g}-\phi_{a})\nonumber\\
 	&\sin(\frac{\omega _a \Delta L}{2v_g}).\label{eq:deltatheta}
 \end{align}
where $\Delta L= L_2-L_1$, $v_g$ denotes the group velocity of spin wave. If $k\Delta L/2=\frac{\pi}{2}$, in the absence of the axion, the two spin waves cancel exactly. Under axion phase modulation at $\omega_a$, the phase shift away from destructive interference yields a magnetization oscillating at $\omega\pm\omega_a$. The magnetization oscillation from the axion-induced phase shift is given by:
\begin{align}
	\Delta M &=2A\cos\omega t \cos \frac{\Delta \theta}{2}\\
	&\approx 2|\chi_+ h_+|\cos(\omega t)\sin \bigl[\frac{g_{ae}a_0}{2m_{e}}v_{az}\nonumber\\
	&\sin(\omega_a t-\frac{\omega_a (L_1+L_2)}{2v_g}-\phi_{a})\sin(\frac{\omega _a \Delta L}{2v_g})\Bigr]\\
	&=2|\frac{\gamma M_0}{\omega_a+i\alpha \omega}|| h_+|\frac{g_{ae}a_0}{2m_{e}}v_{az}\sin(\frac{\omega _a \Delta L}{2v_g})\nonumber\\
	&\frac{1}{2}\bigl[\sin((\omega+\omega_a)t-\phi_L)-\sin((\omega-\omega_a)t-\phi_L)\bigl].
\end{align}
where $\phi_L=\frac{\omega_a (L_1+L_2)}{2v_g}+\phi_{a}$. Maximizing the axion modulation requires the destructive interference condition together with $\omega_a=\pi v_g/\Delta L=v_g k$.     
We take the driving frequency to be equal to the spin wave frequency $\omega_r=\omega$ in the above expression, so as to maximize the spin wave amplitude and ease experimental implementation. In view of the extreme weakness of the axion effect, the sine function is approximated by its argument. Details of the derivation are given in the appendix.
 
 \begin{figure}[t]
 	\centering
 	\includegraphics[width=0.5\textwidth]{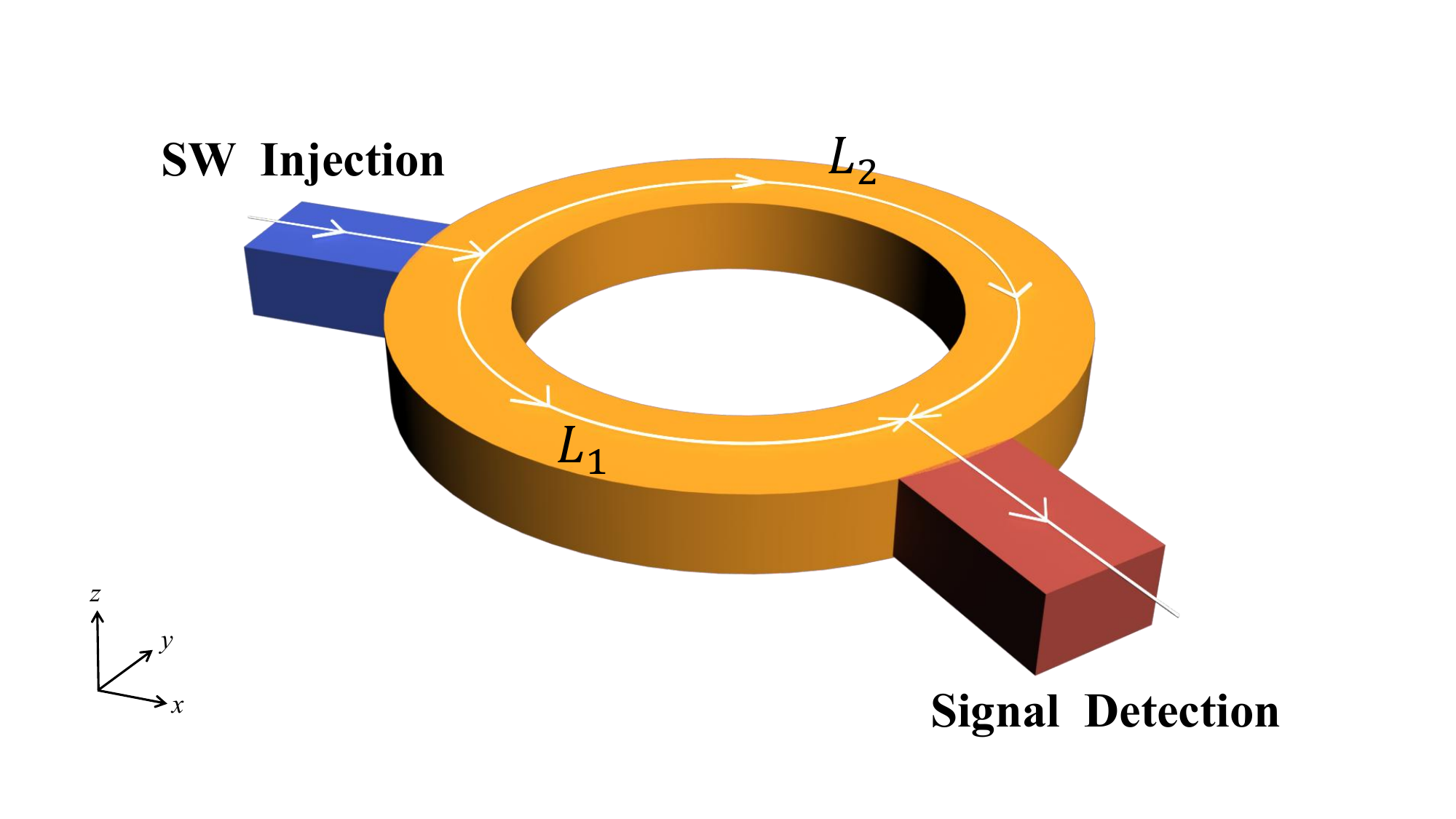}
 	\caption{Conceptual design for axion dark matter detection via a spin wave interferometer. The axion field $a(\bm{r},t)$ induces an effective magnetic field	$\bm{B}_{\mathrm{eff}}$ in the ferromagnet via the axion-electron coupling, which modulates the frequency and phase of spin waves. Two spin waves propagate along paths of different lengths $L_1$ and $L_2$, and then interfere. In the absence of axions, the phase difference satisfies destructive interference, yielding zero output. An axion-induced tiny phase difference breaks the destructive interference, producing a magnetization oscillation with sideband frequencies $\omega\pm\omega_a$. The radiated electromagnetic wave is detected, enabling sensitive detection of axion dark matter. }
\label{con}
 \end{figure}
 
 \textbf{SENSITIVITY} \par
 Radiation from the oscillating magnetization provides a means for axion detection. With a thickness of $5\,\mu m$ and a width of $100\,\mu m$ at the detection site, the spin wave intersection length is $1\,\mu m$, well below the wavelength at 5$\,$GHz for wave number  $k \sim 10^{5}\, m^{-1}$. Therefore, we consider the film volume at the exit of the interference loop to be   $V=1 \,\mu m\times 100 \mu m \times 5 \mu m$, the corresponding radiation power of the magnetization is given by:
 \begin{align}
 	P_\text{out}&=P_{+}+P_{-}\nonumber\\
 	&= \frac{\mu_0 (|\Delta M| V)^2}{12\pi c^3}\Bigl[(\omega+\omega_a)^4 + \lvert\omega-\omega_a\rvert^4\Bigr],
 \end{align}
where $|\Delta M|$ is the amplitude of the magnetization. Subscripts "$+$" and "$-$" mark quantities at frequencies $\omega+\omega_a$ and $\omega-\omega_a$. This radiation can be detected either by amplification with a linear amplifier or by using a single-photon detector integrated with a cavity. The expected rate of emitted photons is
 \begin{align}
 	R_a=\frac{P_\text{out}}{\hbar \omega_s}=\frac{P_+}{\omega+\omega_a}+\frac{P_-}{\omega- \omega_a},
 \end{align}
 where $\omega_s= \omega \pm \omega_a$ \cite{BARBIERI2017135}. We now consider intrinsic magnetization fluctuations $M_n$. The power spectral density of magnetization noise at frequency $\omega_s$ is 
\begin{align}
	(M_n^2)_{\omega_s} = \frac{1}{2\pi V} \frac{k_B T}{\omega_s} \mu''(\omega_s),
\end{align}
where  $\mu''(\omega)$ is the imaginary part of the complex permeability \cite{PhysRevA.79.022118}. $T$ is the temperature of the sample. 

Now we discuss different types of noise.  For the magnetization noise, the radiation spectral density is $S_M $ and the photon rate is $R_M$. A detailed derivation is given in the appendix. The power spectral density of the ambient thermal noise and the photon emission rate of the thermal noise are given by 
\begin{align}
	S_T&=k_B T\\
	R_T&=\bar{n}/\tau_c=\frac{1}{exp[\frac{\hbar \omega_c}{k_B T}]-1}\frac{\omega_c}{Q_c},
\end{align}
where $\tau_c$ represents the cavity decay time, $\omega_c$ and $Q_c$ are the angular resonance frequency and quality factor of the microwave cavity, respectively, and $\bar{n}$ is the average number of thermal photons in an empty cavity \cite{BARBIERI2017135}. 

 Therefore, the signal-to-noise ratios for detection using a linear amplifier and using a single-photon detector \cite{BARBIERI2017135} are correspondingly:
\begin{align}
	&\text{SNR}_{\text{amplifier}}=\frac{P_\text{out}}{P_{\text{M noise}}+P_{\text{T noise}}}\nonumber\\
	&=P_\text{out}\left((S_M^2+S_T^2)_{+}+(S_M^2+S_T^2)_{-  } \right)^{-\frac{1}{2}}\sqrt{\frac{t_m}{\Delta f}}\\
	&\text{SNR}_{\text{photon}}=\frac{R_a}{\sqrt{R_a+R_M+R_T}}\sqrt{\eta t_m},
\end{align}
where $t_m$ is the total measurement time. The instrumental resolution bandwidth reaching the signal bandwidth $\Delta f=\frac{\sqrt{\Delta^2 \omega+\Delta^2 \omega_a}}{2\pi}\approx\frac{\Delta \omega}{2\pi}$, where $\Delta \omega=\omega/Q$ and $\Delta \omega_a=\omega_a/Q_a$ are the bandwidths of the spin wave and the axion signal respectively, with the latter being negligible since quality factor  $Q_a\gg Q$ (e.g., $Q_a\sim10^6$, $Q\sim10^3$). We assume the quantum efficiency of the single-photon detector to be $\eta \approx 1$  as the best estimate. The detectable axion mass ranges from $10^{-8}$ eV to $10^{-6}$ eV, with the lower bound set by the spin wave propagation distance  $L$ and measurement time $t_m$,  and the upper bound by the spin relaxation time \cite{balatsky2023commentaxionmattercouplingmultiferroics}. For axions in this mass range, the corresponding range of the propagation distance difference  $\Delta L$ between the two spin waves is $3\,\mu$m $\sim 300\,\mu$m.
 
\begin{figure}[t]
	\centering
	\includegraphics[width=0.5\textwidth]{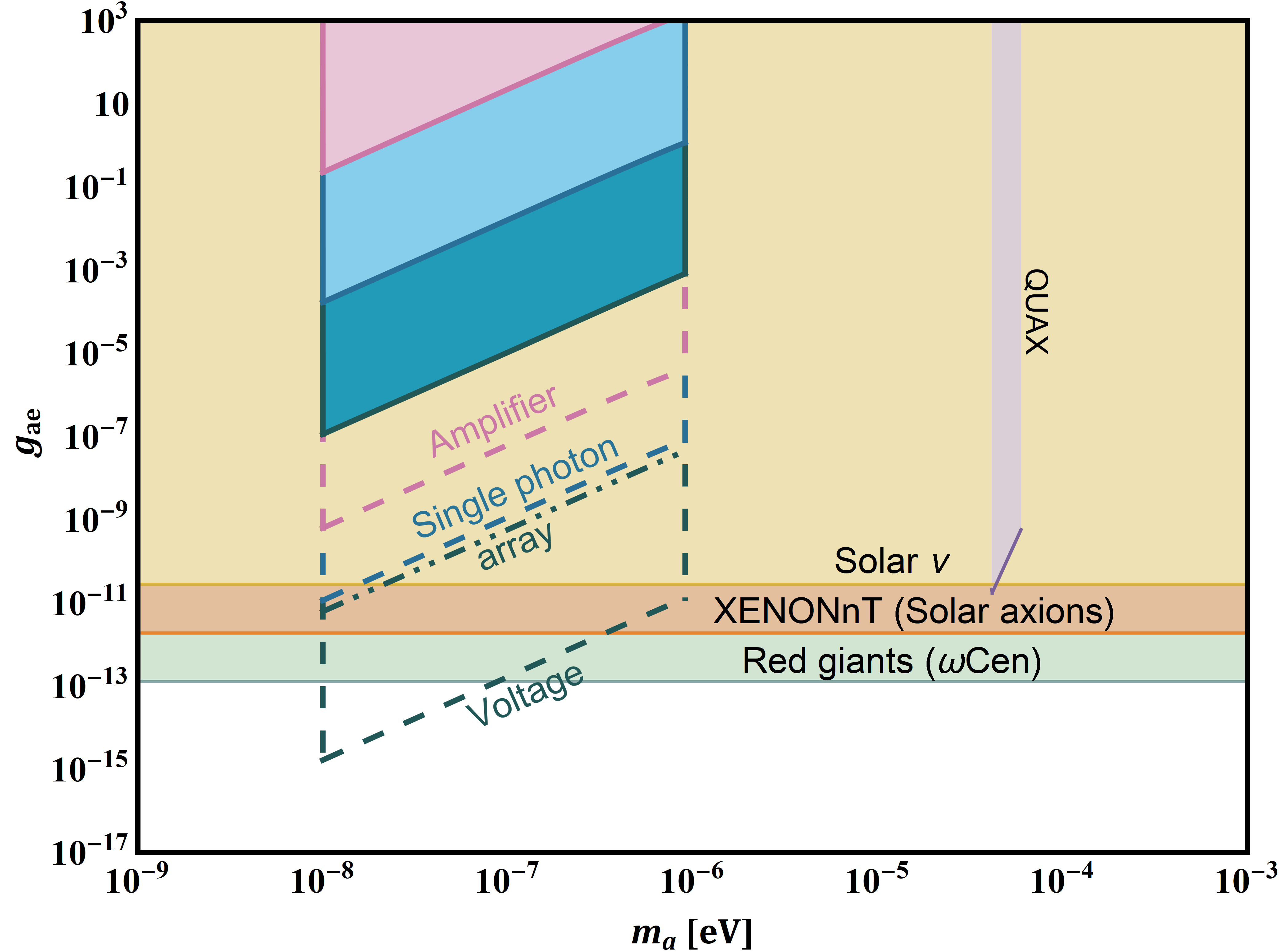}
	\caption {Axion-electron coupling $g_{ae}$ exclusion limits derived from our three detection schemes alongside the projected sensitivity. Pink, light blue, and dark blue exclusion lines denote the linear-amplifier, single-photon-detection, and electrical-detection schemes, respectively. The dashed line corresponds to the exclusion line for the large-volume, low-loss projection scheme. As for the solid line, we take the current realistic value that  $M_s=1.4\times10^5 \,\text{A/m}$, $\alpha=10^{-4}$, $|h|=0.1 \,\text{Oe}$, $H_0 \approx 0.2\,\text{T}$, $T$ = 1 mK, $\omega= \omega_c= 5$ GHz, $Q_c=10^6$, $t_m=1\,\text{year}\sim 3 \times 10^7$ s, $D=8.8\times 10^{-6}\,\text{rad}\cdot \text{m}^2/\text{s} $, $k \sim 10^{5}\, \text{m}^{-1}$.  For the projection scheme, $S=1\,\text{m}\times1\,\text{cm}$, $T_{\text{pro}}=0.1\, $mK, $\alpha = 10^{-7}$.  Under the same conditions as the solid line case, the electrical signal array scheme with $N=10^9$ interferometers is shown as a dash-dot-dot line. Constraints for XENONnT (solar axions), Red giant branch, Solar neutrinos, and the purple QUAX exclusion curve are taken from Refs. \cite{PhysRevLett.131.041003}, \cite{PhysRevD.102.083007}, \cite{PhysRevD.79.107301} and \cite{Crescini_2018,PhysRevLett.124.171801}, respectively. }
\label{sensi}
\end{figure}

In addition to the conventional method of detecting axions via radiation power, the magnetization fluctuations of spin waves can also be combined with Faraday's law to detect the electrical signal induced by spin wave interference \cite{Balynsky:2016uvn}.  At the spin wave intersection, a coil of area \(S = 100\,\mu\text{m} \times 5\,\mu\text{m}\) measures the voltage induced by the magnetic flux variation $\Phi$ through the material cross section.:
\begin{align}
	\varepsilon = -\frac{d\Phi}{dt} = -S \frac{dM}{dt}.
\end{align}
The amplitude of the electromotive force induced by the $10^{-6}$ eV axion is
\begin{align}
	|\varepsilon|=2.4\times 10 ^{-22}  \big(\frac{g_{ae}}{10^{-13}}\big)\text{V} .
\end{align}
Correspondingly, the voltage spectral density arising from magnetization noise is \(V_{\text{M}}\). According to the thermal noise theorem, the thermal noise spectral density of the measurement system is \(V_{\text{noise}}^2 = 4 k_B T R\), where \(R = 50\,\Omega\) is the equivalent noise resistance. Therefore, the signal-to-noise ratio is:
\begin{align}
	\mathrm{SNR}_V = \frac{|\varepsilon|^2}{V_{\text{noise}}^2 + V_{M}^2 }\sqrt{\frac{t_m}{\Delta f}}.
\end{align}
We set \(\text{SNR} = 3\) and obtain the exclusion line. We further extend the scheme to an array of $N=10^9$ spin wave interferometers, which improves the sensitivity by $\sqrt{N}$ due to incoherent noise averaging. The corresponding exclusion curve is plotted in Fig.\ref{sensi}.

For sensitivity projection, the ferromagnetic material cross section(not each loop area) is taken to have a width of $1\,$m and a thickness of $1\,$cm, at a temperature of $T_{\text{pro}}=0.1\,$mK. The Gilbert damping coefficient is set to $\alpha \lesssim 10^{-7}$, rendering magnetization noise negligible, and only thermal noise is considered. The projected exclusion line is derived at SNR = 3, as shown by the dashed curve in Fig.\ref{sensi}, while the solid curve corresponds to the exclusion limit under currently realistic experimental parameters, and the electrical signal array scheme ($N=10^9$) is shown as a dash-dot-dot line.

\textbf{CONCLUSION AND OUTLOOK} \par
In this work, we propose a novel spin wave axion detector. We have theoretically investigated the interaction between axion dark matter and spin waves in ferromagnetic systems. We established a model that describes how to measure an axion-induced effective magnetic field in terms of the ferromagnetic resonance frequency and spin-wave phase, and systematically derived the axion-induced phase-modulation signals in ferromagnetic resonance and spin-wave interferometry.

 The results show that the electrical signal detection scheme offers better sensitivity than the linear amplifier and single-photon schemes, possibly because the electromagnetic induction effect outperforms the radiation effect in this experimental regime. Furthermore, set by the spin wave propagation distance and the spin relaxation time, the axion mass range that can be probed extends from $10^{-8}$ eV to $10^{-6}$ eV. From a projection perspective, a larger cross-sectional area, lower temperature, and smaller Gilbert damping enhance sensitivity, thereby approaching the astrophysical limit.
   
In summary, spin wave phase modulation and interferometry in ferromagnets provide a novel and complementary platform for axion dark matter detection. Future work can be extended to more complex magnetic structures, axion quasiparticles, and optimization of material losses and noise. Since maintaining exact destructive interference is difficult in practice, a common alternative is to introduce a small static phase bias $\epsilon$ away from the exact destructive point.  For the power-dependent readout schemes (linear amplifier and single-photon detector), this makes the signal response linear-dominant ($\propto \epsilon \delta\phi$) rather than quadratic ($\propto \delta\phi^2$), effectively amplifying the weak perturbation.  This approach shares a conceptual similarity with the homodyne detection technique in interferometers such as LIGO\cite{2015,Hild_2009,Fricke_2012}.

Since each detector is rather small ($\sim10^2 \mu$m), an integrated array of such detectors can be assembled to improve sensitivity. More broadly, a distributed array of such interferometers could further map the spatial coherence of the axion field across macroscopic baselines, while also enhancing the detection sensitivity. With ongoing advances in detection and material fabrication techniques, the proposed scheme is expected to be experimentally realizable in the near future, opening new avenues at the interface between condensed matter physics and particle physics. 

\medskip
\textbf{Acknowledgements} \par 
This work is supported by the National Natural Science Foundation of China (No.
12105013, N0. 12447105, No. 5225205, No. T23500), the National Key Research and Development Program of China(No. 2023YFA1406500, J.Z and No. 2021YFA0718700, J.Z), the Fundamental Research Funds for the Central Universities(J.Z) and the Beijing Natural Science Foundation(Z240008, J.Z).05. 

\bibliography{ref.bib}

\clearpage
\appendix
\setcounter{equation}{0}
\newpage
\onecolumngrid

\begin{center}
	\large \textbf{Supplemental material}
\end{center}

\subsection{Spin Wave Coupled to the Background Axion}

The axion couples to fermions via (with \(c = \hbar = 1\))
\begin{align}
\mathcal{L} = \bar{\psi}\left(i\gamma^{\mu}D_{\mu} - m_{f}\right)\psi - g_{af}\frac{\partial_{\mu}a}{2m_{f}}\bar{\psi}\gamma^{5}\gamma^{\mu}\psi,
\end{align}
where \(D_{\mu} = \partial_{\mu} - iqA_{\mu}\) is the covariant derivative, \(g_{af}\) is a dimensionless coupling strength, and \(q\) and \(m_{f}\) are the fermion charge and mass, respectively. \(\psi\) is the Dirac spinor and \(g_{af}\) is a dimensionless coupling strength. The coupling to electrons is limited by \(|g_{ae}| < 3 \times 10^{-11}\) from solar neutrino experiments.  Moreover, \(\gamma^{\mu}\) are Dirac matrices in a particle-hole spinor basis \(\psi = \left(\psi_{f}, \psi_{\bar{f}}\right)\). The Lagrangian yields the equations of motion:
\begin{align}
\left[E + q\varphi - m_{f} + g \grad{a} \cdot \vb{\sigma}\right]\psi_{f} = \left[-g \partial_{t}a + \vb{\sigma} \cdot (\vb{p} - q\vb{A})\right]\psi_{\bar{f}},\\
\left[E + q\varphi + m_{f} + g \grad{a} \cdot \vb{\sigma}\right]\psi_{\bar{f}} = \left[-g \partial_{t}a + \vb{\sigma} \cdot (\vb{p} - q\vb{A})\right]\psi_{f},
\end{align}
where \(g \equiv g_{af}/(2m_{f})\), \(A = (\varphi, \vb{A})\), and \(i \partial_{0}\psi_{f} = E\psi_{f}\), and \(\vb{\sigma}\) is a vector of Pauli matrices. In the nonrelativistic limit, with \(E \approx m_{f}\), and \(q\varphi \ll m_{f}\), these equations give the effective low-energy Hamiltonian:
\begin{align}
H_{af} = \frac{1}{2m_{f}} \left[\vb{\sigma} \cdot (\vb{p} - q\vb{A})\right]^{2} - \frac{g_{af}}{2m_{f}} \left[\grad{a} \cdot \vb{\sigma} + \frac{\partial_{t}a}{m_{f}} \vb{\sigma} \cdot (\vb{p} - q\vb{A})\right] + \mathcal{O}(g_{af}^{2}, \partial^{2}a),
\end{align}
where the left-out terms are either higher order in \(g_{af}\) or in derivatives of \(a\) or in both. The \(\partial_{t}a\) term is expected to be suppressed by \(m_{a}/(2m_{f})\) (which is \(\lesssim 10^{-7}\) for electrons). The so-called "axion wind" coupling \(\sim \grad{a}\) term is relevant for axion dark matter detection.

Now we can look at the effective "axion" part in the material; the axion wind term with \(\sim \grad{a}\) can be the source of the effective magnetic field. We have:
\begin{align}
H_{af} \supset -\frac{g_{af}}{2m_{f}} \grad{a} \cdot \vb{\sigma} = \gamma_{e} \vb{B}_{\rm eff} \cdot \vb{\sigma},
\end{align}
with one extra effective magnetic field added to the spin wave:
\begin{align}
\vb{B}_{\rm eff} = -\frac{g_{af}}{2m_{e}  \gamma_{e}} \grad{a},
\end{align}
where \(\gamma_{e}\) is the gyromagnetic ratio of the electron as \(2\pi \times 28  \, \text{GHz/T}\). Substitute $\gamma_e$ with $\gamma=\mu_0 \gamma_{e}=2.21 \times 10^{5} \ \mathrm{rad/(s\cdot (A/m))}$, the gyromagnetic ratio that multiplies the magnetic field strength $H$ in the LLG equation.

\subsection{Mathematical Derivation}
The modulus of the resultant complex amplitude is computed as:
\begin{align}
	\tilde{A} &= A e^{i\theta_1} + A e^{i\theta_2}, \quad \theta_1 = kL_1+\delta\phi_{1}, \ \theta_2 = kL_2+\delta\phi_{2}. \\
	|\tilde{A}| &= A \sqrt{ \bigl(e^{i\theta_1}+e^{i\theta_2}\bigr)\bigl(e^{-i\theta_1}+e^{-i\theta_2}\bigr) } \nonumber \\
	&= A \sqrt{2 + e^{i(\theta_2-\theta_1)} + e^{-i(\theta_2-\theta_1)}} \nonumber \\
	&= A \sqrt{2 + 2\cos(\theta_2-\theta_1)} \nonumber \\
	&= 2A \left|\cos\frac{\theta_2-\theta_1}{2}\right|.
\end{align}
The phase difference between the two paths is
\begin{align}
	\Delta \theta(L_1,L_2)&= \theta_2-\theta_1\nonumber\\
	&=k\Delta L-\frac{g_{ae}a_0}{2m_{e}}v_{az}[\int_{t-\frac{L_2}{v_g}}^{t}\sin(\phi_a-\omega_a t')dt'-\int_{t-\frac{L_1}{v_g}}^{t}\sin(\phi_a-\omega_a t')dt']\\
	&=k\Delta L-\frac{g_{ae}a_0}{m_{e}}v_{az}\sin(\omega_a t-\frac{\omega_a (L_1+L_2)}{2v_g}-\phi_{a})\sin(\frac{\omega _a \Delta L}{2v_g}).
\end{align}
we take $k\Delta L/2=\frac{\pi}{2}$. 
The magnetization oscillation arising from the phase difference induced by the axion magnetic field can be expressed as:
\begin{align}
	\Delta M &=2A\cos\omega t \cos \frac{\Delta \theta}{2}\\
	&=2|\chi_+ h_+|\cos(\omega t) \sin[\frac{g_{ae}a_0}{2m_{e}}v_{az}\sin(\omega_a t-\frac{\omega_a (L_1+L_2)}{2v_g}-\phi_{a})\sin(\frac{\omega _a \Delta L}{2v_g})]\\
	&\approx 2|\chi_+ h_+|\cos(\omega t) \bigl[\frac{g_{ae}a_0}{2m_{e}}v_{az}\sin(\omega_a t-\frac{\omega_a (L_1+L_2)}{2v_g}-\phi_{a})\sin(\frac{\omega _a \Delta L}{2v_g})\Bigr]\\
	&=2|\frac{\gamma M_0}{\omega_a+i\alpha \omega}|| h_+|\frac{g_{ae}a_0}{2m_{e}}v_{az}\sin(\frac{\omega _a \Delta L}{2v_g})
	\frac{1}{2}\bigl[\sin((\omega+\omega_a)t-\phi_L)-\sin((\omega-\omega_a)t-\phi_L)\Bigr]
\end{align}
 As the axion effect ($g_{ae}$) is extremely weak,  $\sin[\frac{g_{ae}a_0}{2m_{e}}v_{az}\sin(\omega_a t-\frac{\omega_a (L_1+L_2)}{2v_g}-\phi_{a})\sin(\frac{\omega _a \Delta L}{2v_g})]$ is  approximated by $\frac{g_{ae}a_0}{2m_{e}}v_{az}\sin(\omega_a t-\frac{\omega_a (L_1+L_2)}{2v_g}-\phi_{a})\sin(\frac{\omega _a \Delta L}{2v_g})$ in this derivation.
 
 For general magnetic dipole radiation, the radiation power can be expressed as (SI):
\begin{align}
	P=\frac{\mu_0 \omega_m^4 |m|^2 }{12\pi c^3}
\end{align}
where $\omega_m$  is the frequency of the magnetic moment oscillation and $|m|=|M|V$. As the signal to be detected has two frequencies, both frequencies need to be accounted for in the calculations of both the signal and the magnetization noise. 
     \begin{align}
    	P_\text{out}&=P_{+}+P_{-}= \frac{\mu_0 (|\Delta M| V)^2}{12\pi c^3}\Bigl[(\omega+\omega_a)^4 + \lvert\omega-\omega_a\rvert^4\Bigr]\\
    	S_{M_+}&=\frac{\mu_0 M_n^2 V^2}{12\pi c^3}(\omega+\omega_a)^4\\
    	S_{M_-}&=\frac{\mu_0 M_n^2 V^2}{12\pi c^3}(\omega-\omega_a)^4
    \end{align}
where $S_{M_+}$ and $S_{M_-}$ denote the radiation power spectral densities of magnetization noise at frequencies $\omega+\omega_a$ and $\omega-\omega_a$, respectively. The expression for $M_n$ contains $mu''$. Using Eq. \ref{eq:chi} in the main text, we can then obtain:
\begin{align}
\mu''(\omega_s)&=(1+\chi(\omega_s))''=\frac{\alpha \omega \gamma M_s}{(\omega_s-\omega)^2+(\alpha \omega)^2}=\frac{\alpha \omega \gamma M_s}{\omega_a^2+(\alpha \omega)^2}
 \end{align}
Thus, the power spectral density of the magnetization noise is (Natural Units):
\begin{align}
	S_{M_+}&= \frac{V T}{24\pi^2 } \frac{\alpha \omega \gamma M_s}{\omega_a^2+(\alpha \omega)^2} |\omega+\omega_a|^3\\
	S_{M_-}&= \frac{V T}{24\pi^2 } \frac{\alpha \omega \gamma M_s}{\omega_a^2+(\alpha \omega)^2} |\omega-\omega_a|^3
\end{align}
Correspondingly, the rate of radiated photons arising from magnetization noise in the single-photon detection scheme is (Natural Units)
\begin{align} 
	R_M &= \left(\frac{S_{M_+}}{\omega+\omega_a}+\frac{S_{M_-}}{\omega-\omega_a}\right)\frac{\Delta\omega}{2\pi}
\end{align}

For the electrical signal detection method, the amplitude of the signal voltage is
\begin{align}
	|\varepsilon| &= |-\frac{d\phi}{dt} |= |-S \frac{dM}{dt}|\\
	&=S|\chi_+ h_+|\frac{g_{ae}a_0}{2m_{e}}v_{az}\sqrt{\frac{(\omega+\omega_a)^2}{2}+\frac{(\omega-\omega_a)^2}{2}}
\end{align}
 The above formula gives the root-mean-square (RMS) voltage. The two frequencies are different and not integer multiples of each other. They are treated as orthogonal signals in the RMS calculation. The total RMS voltage is equal to the square root of the sum of the squares of the RMS values of each frequency component.
 
 Correspondingly, the voltage arising from magnetization noise at the signal frequency is
\begin{align}
 	V^2_M&=(M^2_n)_{\omega _s}\omega_s^2 S^2\nonumber\\
 	&=\frac{k_B T}{2\pi V}\mu''(\omega_s)S^2\sqrt{\frac{(\omega+\omega_a)^2}{2}+\frac{(\omega-\omega_a)^2}{2}}
\end{align}
\subsection{Path length condition}
Combining the destructive interference condition $k\Delta L/2=\frac{\pi}{2}$ with the condition for maximizing the axion effect $\frac{\omega _a \Delta L}{2v_g}=\frac{\pi}{2}$ yields:
     \begin{align}
     	\omega_a=\pi\frac{v_g}{\Delta L}=v_g k
     \end{align}   
At a spin wave speed of 1500$\, m/s$ and with the propagation distance limited to hundreds of micrometers, the minimum detectable axion mass is on the order of $10^{-8}\,$eV, while for $m_a=10^{-8}\,$eV, $\Delta L \approx 300 \mu m$. Consequently, for the interferometric arms, waveguides with arm lengths on the order of hundreds of micrometers, widths of several tens of micrometers, and thicknesses of a few micrometers may be selected. In the present work, a width of 100 $\mu m$ and a thickness of 5 $\mu m$ are assumed. Provided that the wave transmission remains unaffected, a judicious increase in the material width and thickness can enhance detection sensitivity.  In the present work, we employ YIG as the ferromagnetic material and take the exchange stiffness constant to be $D=8.8\times 10^{-6}\,\text{rad}\cdot m^2/s $, and the static field to be $H_0 \approx 0.2 \text{T}$.
\end{document}